\documentclass[pdflatex,sn-basic, iicol]{sn-jnl}

\usepackage{lineno}
\usepackage{subfig}

\jyear{2022}%

\theoremstyle{thmstyleone}%
\theoremstyle{thmstyletwo}%

\theoremstyle{thmstylethree}%

\newcommand{\bs}[1]{\boldsymbol{#1}}
\newcommand{\mdif}[1]{\dfrac{D #1}{Dt}}
\newcommand{\pdif}[2]{\dfrac{\partial #1}{\partial #2}}

\begin{document}

\title[Topology optimization of soft bodies]{Topology optimization of locomoting soft bodies using material point method}


\author[1]{\fnm{Yuki} \sur{Sato}}\nomail

\author[1]{\fnm{Hiroki} \sur{Kobayashi}}\nomail

\author[1]{\fnm{Changyoung} \sur{Yuhn}}\nomail

\author[1]{\fnm{Atsushi} \sur{Kawamoto}}\nomail

\author*[1]{\fnm{Tsuyoshi} \sur{Nomura}}\email{nomu2@mosk.tytlabs.co.jp}

\author[1]{\fnm{Noboru} \sur{Kikuchi}}\nomail
\equalcont{Present address: \orgname{Toyota Physical and Chemical Research Institute}, \orgaddress{41-1, Yokomichi, Nagakute, \city{Aichi} \postcode{480-1192}, \country{Japan}}}

\affil[1]{\orgname{Toyota Central R\&D Labs., Inc.}, \orgaddress{41-1, Yokomichi, Nagakute, \city{Aichi} \postcode{480-1192}, \country{Japan}}}


\abstract{Topology optimization methods have widely been used in various industries, owing to their potential for providing promising design candidates for mechanical devices.
However, their applications are usually limited to the objects which do not move significantly due to the difficulty in computationally efficient handling of the contact and interactions among multiple structures or with boundaries by conventionally used simulation techniques.
In the present study, we propose a topology optimization method for moving objects incorporating the material point method, which is often used to simulate the motion of objects in the field of computer graphics.
Several numerical experiments demonstrate the effectiveness and the utility of the proposed method.}




\maketitle

\section{Introduction}\label{sec:intro}
Topology optimization methods have successfully been used in many industries, such as automotive industries, owing to their potential for obtaining promising design candidates at the conceptual design stage.
The key concept of topology optimization techniques is the replacement of structural optimization with a material distribution problem, which allows the topological changes in addition to changes in shapes during optimization procedures.
After a pioneering work by \cite{bendsoe1988generating}, topology optimization techniques have been applied to various design problems, including multi-physics problems~\citep{sigmund2001design, dilgen2018density} and manufacturing-oriented design problems~\citep{liu2018current}.
Topology optimization techniques have also been applied to dynamical design problems, such as linkage mechanisms~\citep{han2017topology, han2021topology} and gear-linkage mechanisms~\citep{yim2019topology}. 
However, topology optimization methods have rarely been applied to design the structure of locomoting soft bodies, which involve large deformations of materials themselves. 
It would be due to the difficulty in handling the contact and interaction among multiple structures or with boundaries considering the geometrical non-linearity in a computationally efficient manner by conventionally used simulation techniques represented by the finite element method (FEM).

Handling locomoting soft bodies is essential in the field of soft robotics for developing deformable robots.
This field has recently attracted much attention due to the relation to multidisciplinary fields, covering electronics, materials science, computer science, and biomechanics~\citep{laschi2014soft, whitesides2018soft}, which implies that the field has the potential for providing various applications.
In the literature, several previous studies proposed controller optimization of locomoting soft robots~\citep{george2018control, kim2021review, hu2019difftaichi, bhatia2021evolution}.
Additionally, \cite{cheney2014unshackling} performed design optimization of locomoting soft robots using the compositional pattern-producing network (CPPN).
\cite{bhatia2021evolution} also performed design
optimization of locomoting soft robots based on genetic algorithms and Bayesian optimization in addition to CPPN.
However, since the number of design variables this method can efficiently handle is relatively small, the degree of design freedom is limited.
Therefore, there is room for improving the performance of soft robots by further optimizing their structures.
Furthermore, this opens up a new possibility of finding a novel soft mechanism by computational design.

For topology optimization of locomoting soft bodies, numerical simulation techniques, which are computationally efficient in handling the dynamics of soft bodies, are required.
One of the possible candidates is the material point methods (MPMs)~\citep{sulsky1994particle, jiang2016material}, which were originally introduced in computational mechanics and are now widely used to efficiently compute the motion of moving objects in the field of computer graphics~\citep{stomakhin2013material, stomakhin2014augmented}.
MPMs are the hybrid Lagrangian--Eulerian techniques and have advantages over conventional Lagrangian or Eulerian methods.  
Compared with FEMs, which are conventionally used in topology optimization, the MPM can easily handle large deformation, collision, and contact, owing to the hybrid Lagrangian--Eulerian approach.

When using the FEMs, the large deformation simulations often cause numerical instabilities due to excessive distortion of meshes having low stiffness~\citep{wang2014interpolation}.
On the other hand, MPMs easily simulate large deformation because they are mesh-free approaches, tracking the deformation using particles in the Lagrangian description. 
Furthermore, MPMs also simulate the collision and contact relatively easily by calculating the interaction on grids in the Eulerian description. 
Thus, the hybrid Lagrangian--Eulerian approach makes it easier to simulate large deformation, collision, and contact.

Recently, a topology optimization method incorporating MPMs has been proposed to optimize elastic materials with large static deformation~\citep{li2021lagrangian}.
Another technique to be required for topology optimization of locomoting soft bodies is the topology optimization for time-domain problems.
\cite{wang2020space} proposed space--time topology optimization where the time-dependent material distribution is optimized for additive manufacturing.
\cite{nomura2007structural} and \cite{yaji2018large} formulated the optimization problems in the time domain where the forward and sensitivity analyses were performed in a relevant time interval.
In the present study, we propose a topology optimization method for locomoting soft bodies incorporating MPMs, formulating the optimization problem in the time domain.
The proposed method incorporates a material representation scheme for topology optimization with MPMs to optimize locomoting soft bodies as material distribution, with a formulation of a density filtering in the framework of MPMs for obtaining smoothed structures.
The contributions of the present study are the followings:
\begin{itemize}
    \item We incorporated the MPM, originally introduced in computational mechanics but now widely used in computer graphics, with structural optimization methods, especially topology optimization methods.
    This incorporation enables topology optimization of locomoting soft bodies, which might otherwise be challenging for conventional schemes.
    \item We introduced topology optimization, i.e., material distribution optimization, into the design optimization of locomoting soft bodies.
    This makes the degree of the design freedom much higher compared to the existing approaches for designing soft bodies.
    \item 
    We fully exploited the framework of DiffTaichi~\citep{hu2019difftaichi}, especially for the sensitivity calculation of many design variables. 
    The utilization of the automatic differentiation (AD) is essential for dynamic design problems, which involve complicated situations such as contacts. 
\end{itemize}

\section{Formulations}\label{sec:method}

\subsection{Governing equation} \label{sec:mpm}
Firstly, the conservation law of the mass and momentum reads
\begin{align}
    & \mdif{\rho} + \rho \nabla \cdot \bs{v} = 0, \label{eq:mass} \\
    & \rho \mdif{\bs{v}} = \nabla \cdot \bs{\sigma} + \rho \bs{g}, \label{eq:momentum}
\end{align}
where $\rho$ is the density, $t$ is the time, $\bs{v}$ is the velocity, $\bs{\sigma}$ is the Cauchy stress tensor, and $\bs{g}$ is the gravity acceleration.
The operator $D/Dt$ represents the material derivative.
Let $\bs{F}$ denote the deformation gradient, then the Cauchy stress is given as
\begin{equation}
    \bs{\sigma} = \frac{1}{J} \pdif{\varPsi}{\bs{F}} \bs{F}^\top, \label{eq:Cauchy}
\end{equation}
where $J=\mathrm{det}(\bs{F})$ and $\varPsi$ is the potential energy.
Assuming the neo-Hookean material constitution, the potential energy $\varPsi$ is given as follows:
\begin{align}
    \varPsi(\bs{F}) := \dfrac{\mu}{2} \left( \mathrm{tr}(\bs{F}^\top \bs{F}) - d \right) - \mu \log (J) + \dfrac{\lambda}{2} \log^2 (J),
\end{align}
where $d$ is the spatial dimensions.
These governing equations are solved using the material point method (MPM).
Specifically, in the present study, we use the moving least square material point method (MLS-MPM)~\citep{hu2018moving}, which incorporates the moving least square discretization of the governing equation with the MPM.

\subsection{Material representation in material point method} \label{sec:matrep}
We now describe the material representation in the material point method.
Since the governing equations include a self-weight load, i.e. the gravity acceleration, certain treatments in material interpolation schemes are required for dealing with it.
Several previous studies~\citep{bruyneel2005note, kumar2022topology} proposed novel interpolation schemes of the mass density so that numerical instabilities at low densities could be avoided and the non-monotonous behaviors of objective functions could be reduced.
In the present study, since the objective function is originally non-monotonous, different from structural compliance, the SIMP interpolation does not contribute to penalizing the intermediate values of design variables, i.e., grayscales.
Thus, we used a linear interpolation scheme both for stiffness and mass density, which avoided the ratio of the gravity force to stiffness from diverging and made it easier to handle self-weight problems.
Let $\gamma_p$ for $p=1, \ldots, N_p$ denote the fictitious material density of the $p$-th particle in the design domain where $N_p$ is the number of particles in the domain.
Then, the density and the Lam\'{e}'s constants of the $p$-th particle are linearly interpolated using $\gamma_p$ as follows:
\begin{align}
    \rho_p & = \rho \left( ( 1 - \varepsilon) \gamma_p + \varepsilon \right), \\
    \lambda_p & = \lambda \left( ( 1 - \varepsilon) \gamma_p + \varepsilon \right), \\
    \mu_p & = \mu \left( ( 1 - \varepsilon) \gamma_p + \varepsilon \right),
\end{align}
where $\lambda$ and $\mu$ are the Lam\'{e}'s constants of the material, and $\varepsilon$ is a small constant, which is introduced to avoid the singularity.

\begin{figure*}[t]
    \centering
    \includegraphics[width=0.7\textwidth]{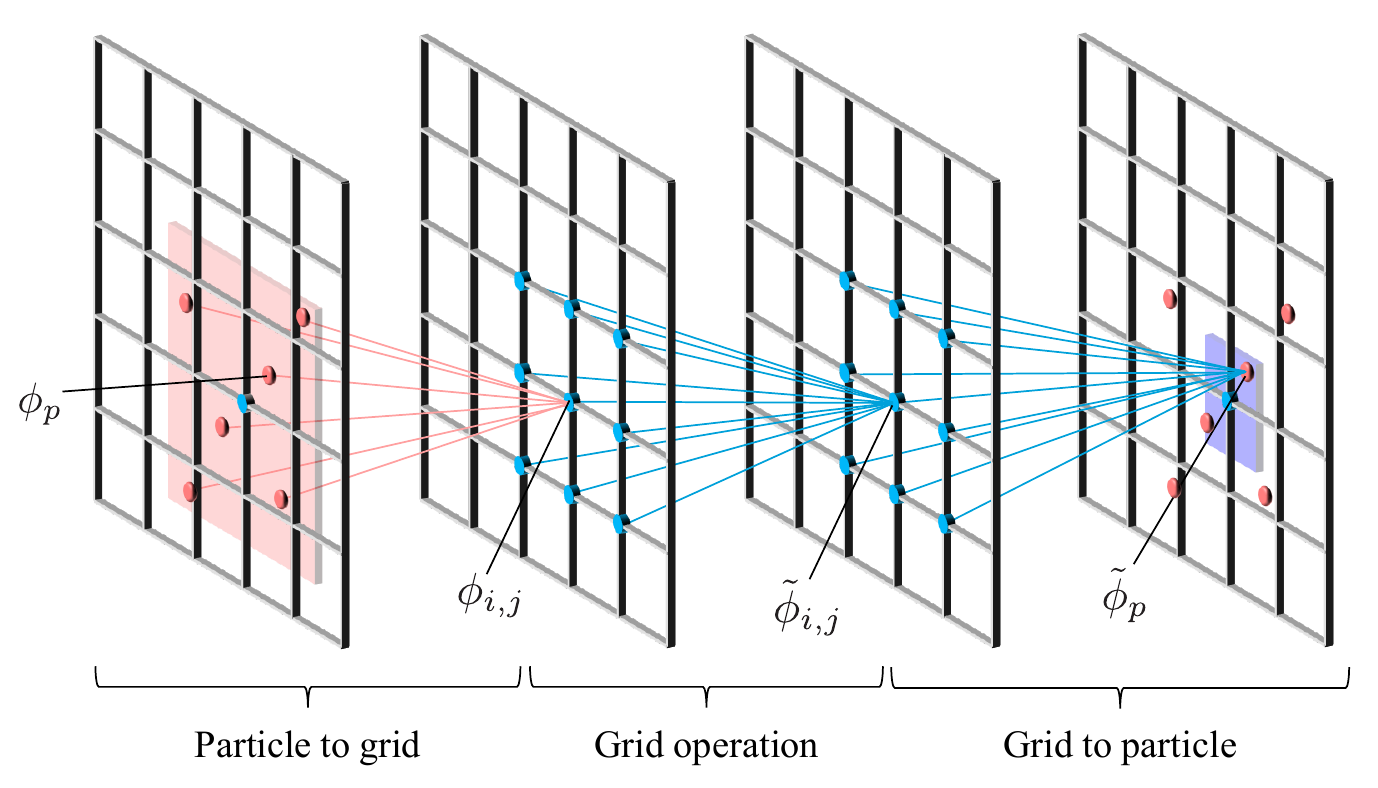}
    \caption{Schematic of the density filtering. In the particle-to-grid process, design variables of particles in the red region are transferred to the $(i, j)$ grid (blue node). In the grid operation, smoothed grid design variables are calculated as the convolution of the grid design variables. In the grid-to-particle process, smoothed grid design variables at $(i+k, j+l)$ grids for $k, l \in \{-1, 0, 1\}$ ($3 \times 3$ blue nodes) are transferred to smoothed design variables of particles in the blue region.}
    \label{fig:grid_filter}
\end{figure*}
\subsection{Heaviside projection with density filtering}\label{sec:filter}
To ensure the smoothness of the structures, topology optimization methods usually employ a certain filtering technique.
In the present study, we introduce a filtering technique in the scheme of the material point method.
Let $\phi_p$ denote the auxiliary design variables assigned to the $p$-th particle to be designed.
These design variables are spatially filtered as shown in Fig.~\ref{fig:grid_filter} and projected onto the fictitious particle densities $\gamma_p$, as follows:
\begin{enumerate}
    \item \textbf{Particle to grid:} Transfer the particle design variables $\phi_p$ to grid design variables using the same framework as the MLS-MPM. 
    The transferred grid design variables are then normalized by the sum of the weights of each particle to each grid.
    \item \textbf{Grid operation:} Take the weighted sum of the grid design variables in the neighboring grid points.
    For two-dimensional problems, for example, let $\phi_{i,j}$ and $\tilde{\phi}_{i,j}$ respectively denote the grid design variables and smoothed grid design variables on the $(i, j)$ grid.
    Then the smoothed grid design variables are calculated as the convolution of the grid design variables, as follows:
    \begin{align}
        \tilde{\phi}_{i,j} =& w_0\phi_{i,j} + w_1\phi_{i-1,j} + w_1\phi_{i+1,j} \nonumber \\
        &+ w_1\phi_{i,j-1} + w_1\phi_{i,j+1} + w_2\phi_{i-1,j-1} \nonumber \\
        &+ w_2\phi_{i+1,j-1} + w_2\phi_{i-1,j+1} + w_2\phi_{i+1,j+1},
    \end{align}
    where $w_0, w_1, w_2$ are weighting coefficients satisfying $w_0 + 4w_1 + 4w_2 = 1$.
    These parameters $w_0$, $w_1$, and $w_2$ control the degree of smoothness of filtered variables.
    \item \textbf{Grid to particle:} Transfer the smoothed grid design variables to the smoothed particle design variables $\tilde{\phi}_p$ using the same framework as the MLS-MPM.
    \item \textbf{Heaviside projection:} Project the smoothed particle design variables $\tilde{\phi}_p$ onto the fictitious particle densities using the smoothed Heaviside function as follows:
    \begin{equation}
        \gamma_p = \dfrac{1}{2} \left( \dfrac{\tanh{(\beta \tilde{\phi}_p)}}{\tanh{(\beta})} + 1\right),
    \end{equation}
    where $\beta$ is a parameter to control the curvature of the smoothed Heaviside function.
\end{enumerate}

\subsection{Problem statement}\label{sec:problem}
In the present study, we focus on the problem of designing soft robots that move toward the designated direction as far as possible.
Consider that a soft robot to be designed consists of the fixed design domain, the actuation domain, and the non-design domain.
In the fixed design domain, the material distribution is optimized, while in the other domains, the structure is unchanged during the optimization process.
In the actuation domain, the prescribed stress $\bs{\sigma}_\mathrm{act}$ is added to the Cauchy stress in Eq.~\eqref{eq:Cauchy}.
The prescribed stress $\bs{\sigma}_\mathrm{act}$ is given as
\begin{equation}
    \bs{\sigma}_\mathrm{act} (t) = \sum_{k=1}^{N_a} \left( a_k \sin(\omega_k t + \pi\theta_k) + b_k \right) \bs{F}(t) \bs{S}_k \bs{F}(t)^\top,
\end{equation}
where $\bs{S}_k$ is the second Piola-–Kirchhoff stress caused by the $k$-th actuator, $N_a$ is the number of actuators, $a_k$ is the amplitude, $\theta_k$ is the phase, $b_k$ is the bias, and $\omega_k$ is angular frequency.

The objective is to maximize the distance that an object travels, which is measured by the mass center in the designated direction $\bs{e}$ at the end of the prescribed time interval as follows:
\begin{align}
    \mathcal{L} := \dfrac{\int_{\Omega_N (T)} \bs{x} \cdot \bs{e} ~ d\bs{x}}{ \int_{\Omega_N (T)} d\bs{x} },
\end{align}
where $\Omega_N (T)$ represents the union of the non-design and actuation domains at predetermined time $T$.
Here, we define the objective function within these domains so that the domain of integration should be unchanged during the optimization procedure.
The design variables are $\bs{\phi} = \left[ \phi_1, \ldots, \phi_{N_p} \right]^\top$ for material configurations and the input parameters to actuators $\bs{a} = \left[ a_1, \ldots, a_{N_a} \right]^\top$, $\bs{\theta} = \left[ \theta_1, \ldots, \theta_{N_a} \right]^\top$, and $\bs{b} = \left[ b_1, \ldots, b_{N_a} \right]^\top$.
All design variables are bounded within $[-1, 1]$.
The optimization problem is then formulated in the minimization form as
\begin{subequations}
\begin{alignat}{2}
    \min_{\bs{\phi}, \bs{a}, \bs{\theta}, \bs{b}}& \quad && - \mathcal{L}, \\
    \text{subject to} & \quad && \bs{\phi} \in [-1, 1]^{N_p}, \\
    & \quad && \bs{a}, \bs{\theta}, \bs{b} \in [-1, 1]^{N_a},
\end{alignat}
\end{subequations}
in which Eqs.~\eqref{eq:mass} and \eqref{eq:momentum} are nested.

\section{Numerical experiments}\label{sec:result}
In the following, we provide two numerical experiments.
The proposed method was implemented by Python and specifically, the forward and backward analysis using the MPM was implemented by Taichi~\citep{hu2019taichi,hu2019difftaichi}, which can be called in Python environment.
Taichi is an open-source programming language for high-performance numerical computation, and provides a scheme for automatic differentiation which we used for sensitivity calculations.
Lam\'{e}'s constants were respectively set to $\lambda=10$ and $\mu=10$, and the small constant $\varepsilon$ was set to $10^{-3}$.
The density was set to $\rho=1$.
The number of actuators $N_a$ was set to $1$, and the angular frequency of the actuators $\omega_1$ was set to $30$.
The prescribed second Piola-Kirchhoff stress was given as $S_1 = 2I$ where $I$ is the identity matrix.
The weighting coefficients for density filtering were set to $w_0=0.64$, $w_1=0.08$, and $w_2=0.01$.
These parameters $w_0$, $w_1$, and $w_2$ control the degree of smoothness of filtered variables and were determined empirically through numerical experiments.
The number of grids in the MPM was $128 \times 128$, and the particles were initially aligned in a grid-like manner with a doubled resolution to the grids.
The forward simulation was performed $5000$ time steps with the time interval $2.0 \times 10^{-4}$~s.
The optimization was performed using the method of moving asymptotes~\citep{svanberg1987method}.
The parameter $\beta$ in the smoothed Heaviside function was initially set to $0.5$ and was doubled when the optimization process converged, until the parameter reached $16$.
The optimization process continued while the change in the moving average of the objective function values over $10$ iterations was larger than $2.0 \times 10^{-4}$.

\subsection{Walker design}
\begin{figure}
    \centering
    \includegraphics[width=\linewidth]{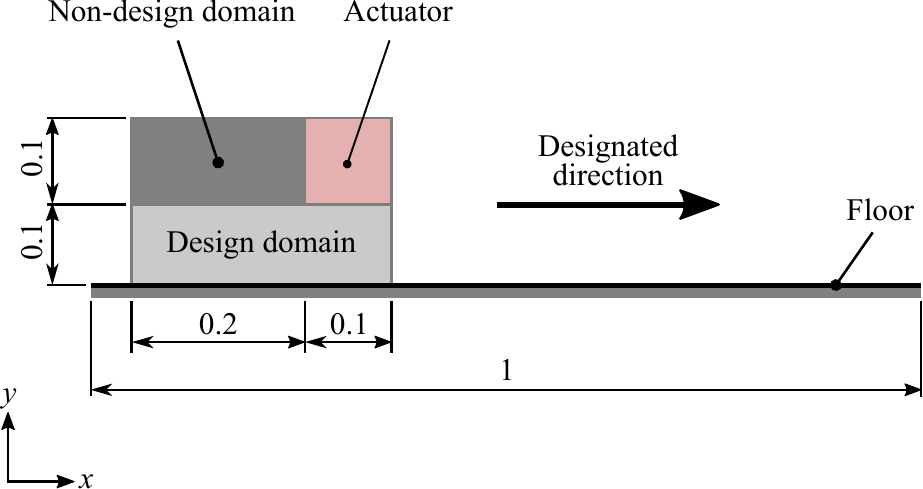}
    \caption{Walker design problem}
    \label{fig:schematic_walker}
\end{figure}

\begin{figure}
    \centering
    \includegraphics[width=\linewidth]{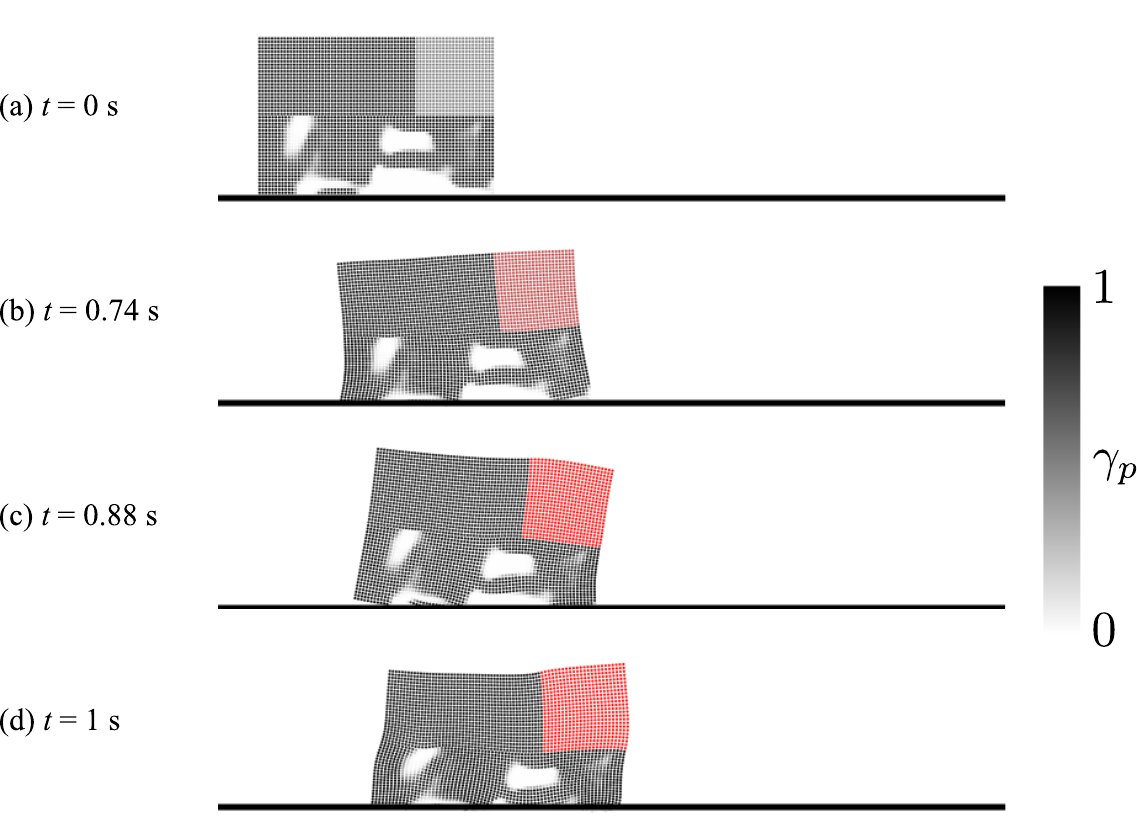}
    \caption{Optimized result of a walker. The grayscale of the design domain represents the value of $\gamma_p$, and the red area represents the positive stress induced by the actuation.}
    \label{fig:opt_result_walker}
\end{figure}

\begin{figure}
    \centering
    \includegraphics[width=\linewidth]{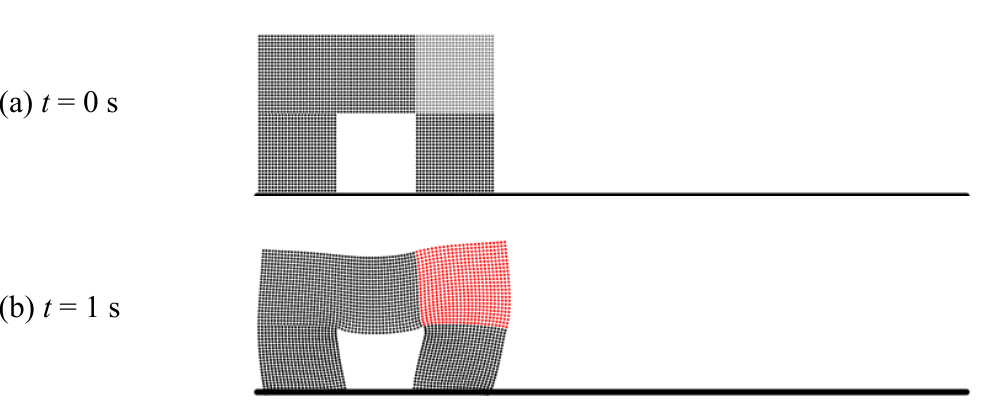}
    \caption{Forward simulations of the reference design of a walker.}
    \label{fig:reference_walker}
\end{figure}

The first problem setting for designing a walking robot is illustrated in Fig.~\ref{fig:schematic_walker}.
The gravity acceleration $\bs{g}$ is set to $-10$ along the $y$-axis.
The optimization result is shown in Fig.~\ref{fig:opt_result_walker}, which exhibits that the obtained structure can walk toward the designated direction, i.e., along the $x$-axis.
The input parameters of the actuator were optimized to $a=1.00$, $\theta=-0.330$, and $b=-0.982$.
In the optimized structure, \textit{legs}-like substructures were formed, which took a step toward the designated right direction as shown in Fig.~\ref{fig:opt_result_walker}~(b) and (c).
To verify the optimized result, we further performed forward simulations of the walker's body exactly represented by particles placed where $\rho \geq 0.3$, which was empirically set.
The post-processed results still exhibit better performance compared to the reference designs shown in Figs.~\ref{fig:reference_walker}.
The objective function values were listed in Table~\ref{table:walker_Lval}.


\begin{table}[t]
	\caption{Objective function values of the walker.}
	\label{table:walker_Lval}
	\centering
	\begin{tabular}{ll}
		\hline
		Condition & Objective function value \\
		\hline \hline
		Optimized & -0.362 \\
		Post-processed & -0.357 \\
		Reference design & -0.208 \\
		\hline
	\end{tabular}
\end{table}

\subsection{Crawler design}
\begin{figure}
    \centering
    \includegraphics[width=\linewidth]{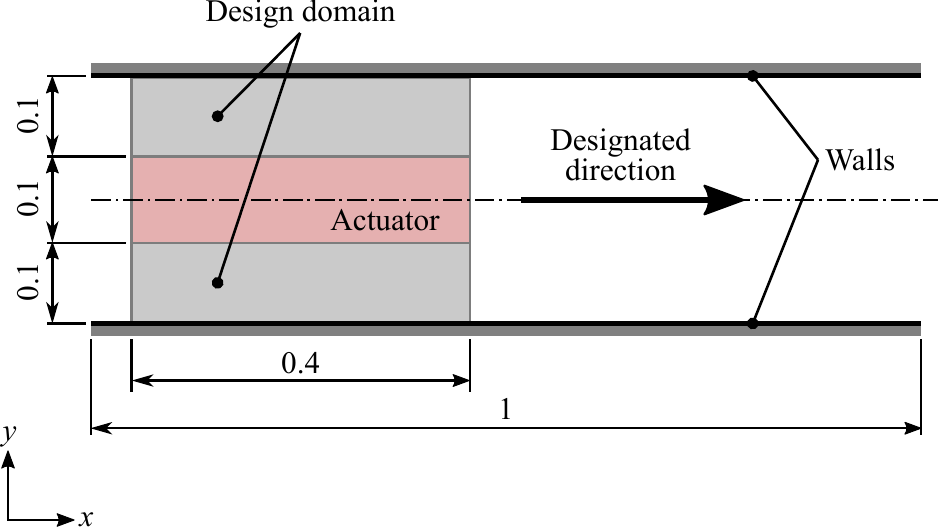}
    \caption{Crawler design problem}
    \label{fig:schematic_crawler}
\end{figure}

\begin{figure}
    \centering
    \includegraphics[width=\linewidth]{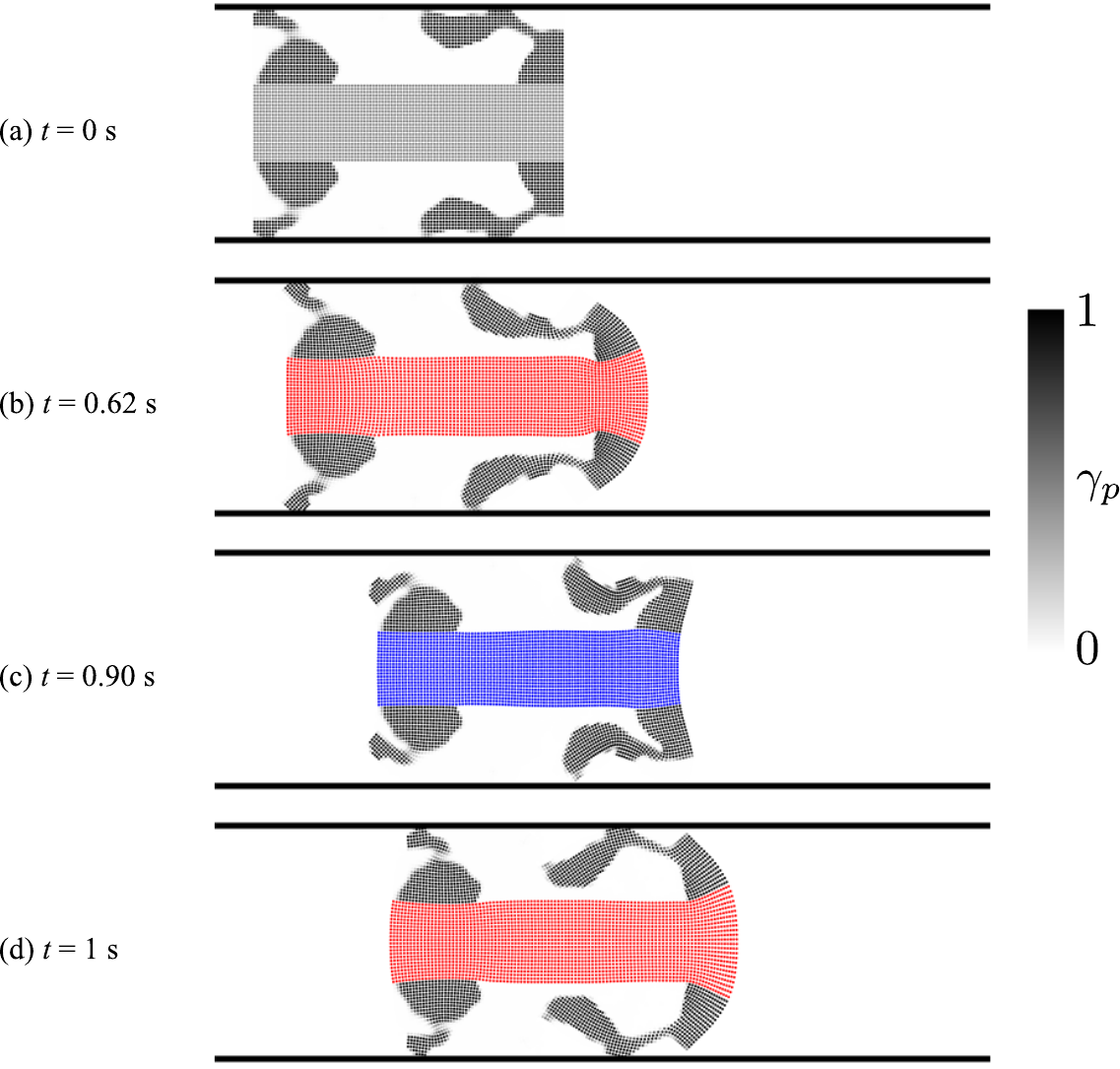}
    \caption{Optimized result of crawler. The grayscale of the design domain represents the value of $\gamma_p$, and the red (blue) area represents the positive (negative) stress induced by the actuation.}
    \label{fig:opt_result_crawler}
\end{figure}

\begin{figure}
    \centering
    \includegraphics[width=\linewidth]{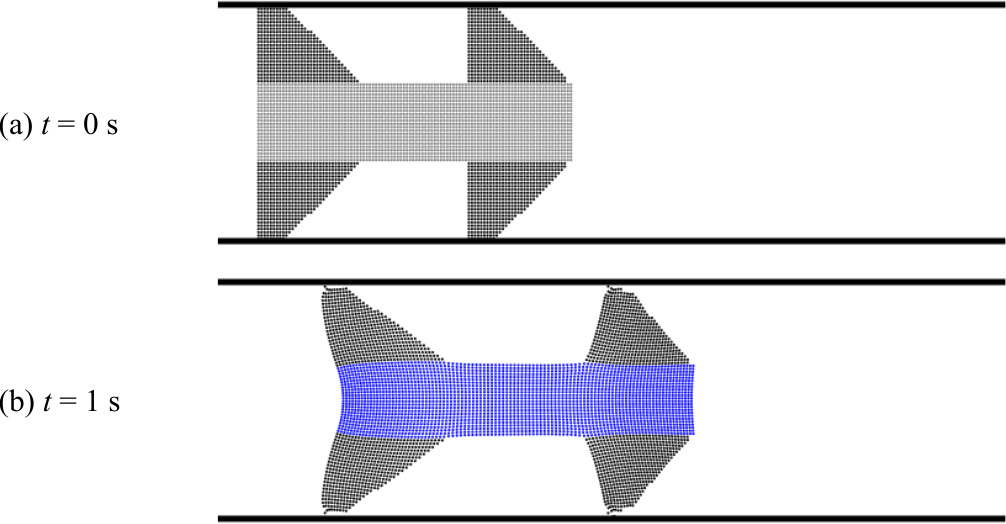}
    \caption{Forward simulations of the reference design of a crawler.}
    \label{fig:reference_crawler}
\end{figure}

The second problem setting for designing a crawling robot is illustrated in Fig.~\ref{fig:schematic_crawler}.
In this problem, no gravity is applied.
Moreover, the grid design variables in the lower half are mapped to those in the upper half just after Step 2 in Sec.~\ref{sec:filter} to keep the design symmetry across the horizontal mid-axis.
Figure~\ref{fig:opt_result_crawler} shows the optimized result of the crawler design. 
The input parameters of the actuator were optimized to $a=0.964$, $\theta=-0.174$, and $b=-0.447$.
As shown in this figure, the optimized structure moves forward by alternatively kicking the walls with the front and rear \textit{legs}.
As a verification of the results, we performed forward simulations in which the crawler's body was exactly represented by particles placed where $\rho \geq 0.3$.
The post-processed results exhibit better performance compared to the reference designs shown in Fig.~\ref{fig:reference_crawler} as listed in Table~\ref{table:crawler_Lval}.


\begin{table}[t]
	\caption{Objective function values of the crawler.}
	\label{table:crawler_Lval}
	\centering
	\begin{tabular}{ll}
		\hline
		Condition & Objective function value \\
		\hline \hline
		Optimized & -0.451 \\
		Post-processed & -0.401 \\
		Reference design & -0.379 \\
		\hline
	\end{tabular}
\end{table}

\section{Conclusion}\label{sec13}
This paper presented a topology optimization for locomoting soft bodies incorporating the material point method.
We proposed a material representation scheme in MPMs and constructed a density filtering technique in the framework of MPMs, which enables us to use MPMs in topology optimization.
The effectiveness of the proposed method was confirmed through numerical experiments.
On the other hand, the present study has limitations on manufacturability due to the material property setting.
Since the dynamics we focus on are geometrically non-linear, the material property and geometry must be set based on the real-world environment.
Albeit such requirements, the present study used numerically convenient settings for simplicity.
We hope to address manufacturability issues in our future research.
We believe that the proposed method will broaden the field of applications of topology optimization for the dynamics of soft bodies, which is related to large deformation, collision, and contact.

\backmatter

\section*{Statements and Declarations}
\bmhead{Conflict of interest}
On behalf of all authors, the corresponding author states that there is no conflict of interest.
\bmhead{Replication of results}
The source code is unavailable due to institutional constraints. However, further algorithm details are available upon request to the authors.
\section*{Acknowledgement}
The manuscript has no funding for this paper or is not applicable.

\bibliography{mybib}


\end{document}